\documentclass[]{aastex631}

\usepackage{hyperref}

\begin{document}

\title{PERISTOLE: PackagE that geneRates tIme delay plotS caused by graviTatiOnaL lEnsing}

\correspondingauthor{T.S.Sachin Venkatesh}
\email{tssachin.venkatesh@gmail.com}

\author[0000-0002-8793-2569]{T.S.Sachin Venkatesh}
\affiliation{Delhi Technological University\\
Delhi 110042, India}

\author{Gaurav Pundir}
\affiliation{Indian Institute of Science Education and Research, Pune\\
Maharashtra 411008, India}



\begin{abstract}

We present PERISTOLE to study the various time delays associated with the pulsar rotation and other general relativistic aspects of binary pulsars. It is made available as an open-source python package which takes some parameters of the double pulsar system as input and outputs the rotational and latitudinal lensing delays along with the geometric and Shapiro delays that arise due to gravitational lensing. This package was intended to provide a way to quickly analyse, evaluate and study the differences between variations of the same systems and also to quantify the consequences that different parameters have over the system. Through this research note, we briefly describe the motivation behind PERISTOLE and showcase its capabilities using the only double pulsar system ever found, J0737-3039.
\end{abstract}


\section*{Introduction}
\subsection*{B\lowercase{inary} P\lowercase{ulsars}}
Binary pulsars have been recorded to have a periodic variation in their pulse arrival time which warrants the use of relativistic corrections to adjust for their timing model, especially for close binaries which include double Pulsar systems, Pulsar-White Dwarf systems and Pulsar-Neutron Star systems, generally noted as binary Pulsar-body systems from now on. Such systems are described to be excellent astrophysical laboratories to test for extreme general relativistic effects and are prime candidates for ushering into the next stage of gravitational wave physics \citep{Taylor1992}. But these double Pulsars and Pulsar-NS systems are an elusive breed and their rarity comes from the condition that they not only have to have survived multiple mass transfer stages, accretion leading to two late stage supernovae processes but also the nomenclature condition that neutron stars can be classified as pulsars only if their radio beams point towards us. This, mounting upon the condition that they have to have near edge-on orbit configurations with respect to the observer for proper analysis further shrinks the sample space available to us for a large scale investigation. 

The orbital motion of a pulsar gives rise to special relativistic aberration affecting the emission direction with respect to its pulsar spin axis \citep{Komesaroff1970}. The longitudinal change gives rise to the longitudinal delay which shifts the arrival time of the pulse while the latitudinal change results in the distortion of the pulse profile. The bending of the radio beam also introduces rotational delay, accentuated at the point of superior conjunction. Additionally, this light bending also splits the pulse profile into two gravitationally lensed images of the source which are defined as dominant and sub-dominant images.  

PERISTOLE aims to act as a mock analysis and graphing tool where the user can simulate different binary Pulsar-body systems to study the aforementioned delays and determine the role that the system parameters play in accentuating or extenuating these delays.

\subsection*{PSR J0737-3039}
There have been various studies on the different types of delays associated with Double Neutron Star (DNS) systems supported by observational evidence but there has been only one study in similar standing until now for double pulsar systems \citep{Lorimer2005}, simply due to the fact that only one such system has ever been found, PSR J0737-3039. We now list some of the system parameters relevant to this study. This system's primary pulsar, `A' has a mass of 1.337\(M_\odot\) and the companion, `B' has a mass of 1.25\(M_\odot\). `A' is a milli-second pulsar with a period of 22.7 ms while the companion has a period of 2.773 sec. The orbital semi-major axis of the system is 8.784 $\times$ $10^{8}$ m with an eccentricity of 0.0878 and its longitude of perisastron being 74\textdegree. This system has a near edge-on configuration which aids in the accurate analysis of the different delays associated with the system. 

\section*{Functionality}
\begin{figure}
    \includegraphics[width=.5\textwidth]{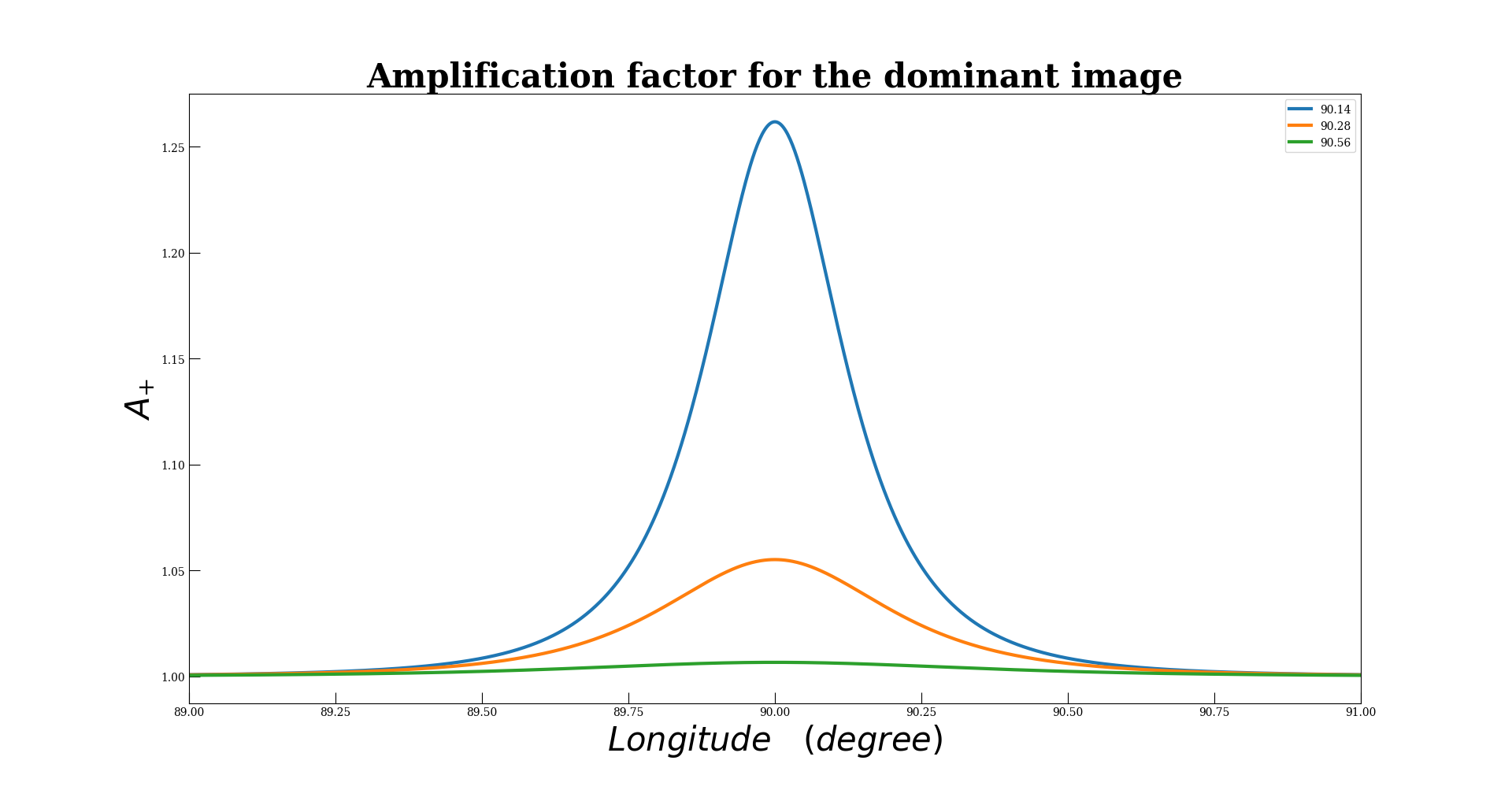}
    \includegraphics[width=.5\textwidth]{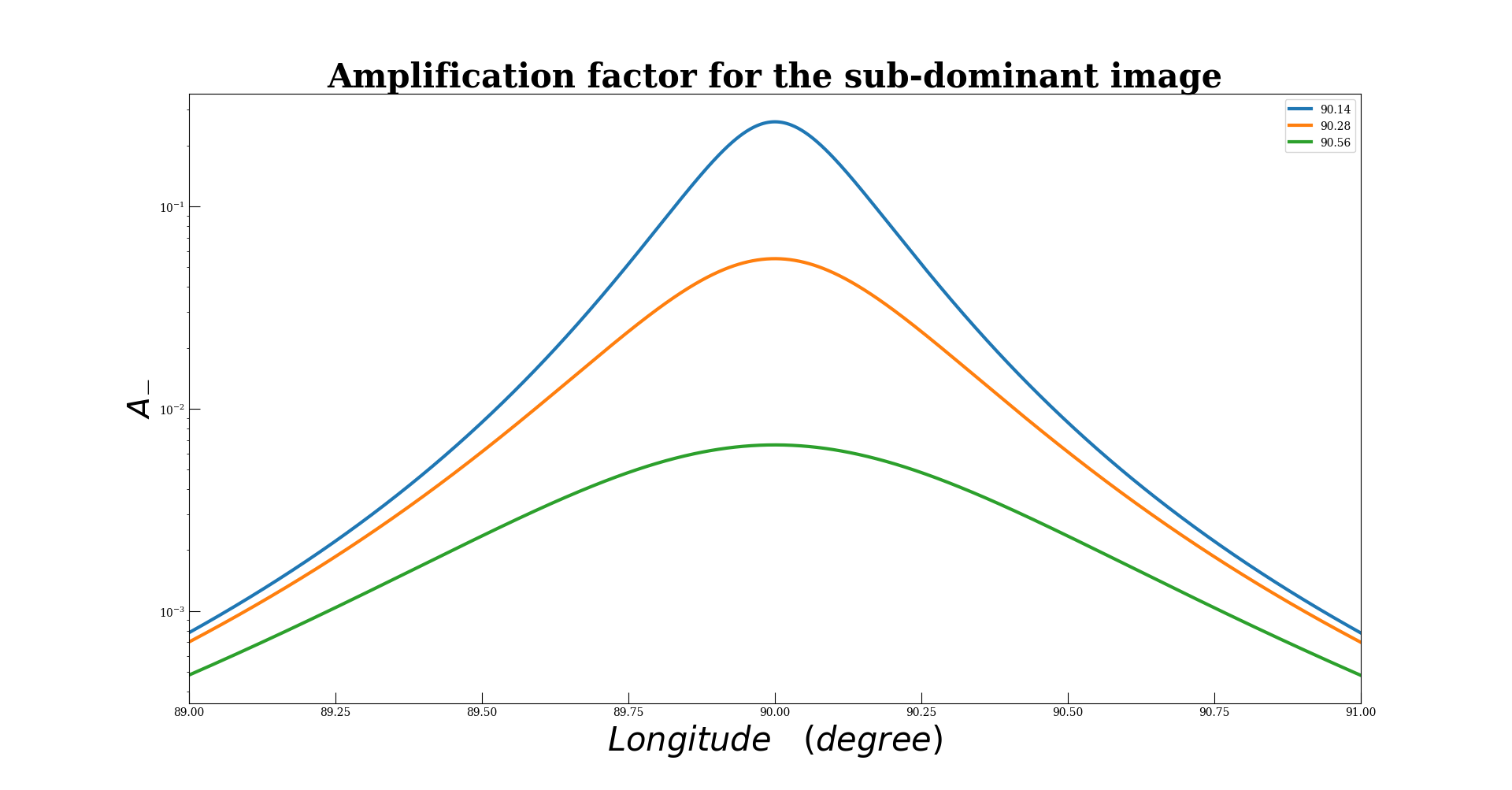}
    \newline
    \includegraphics[width=.5\textwidth]{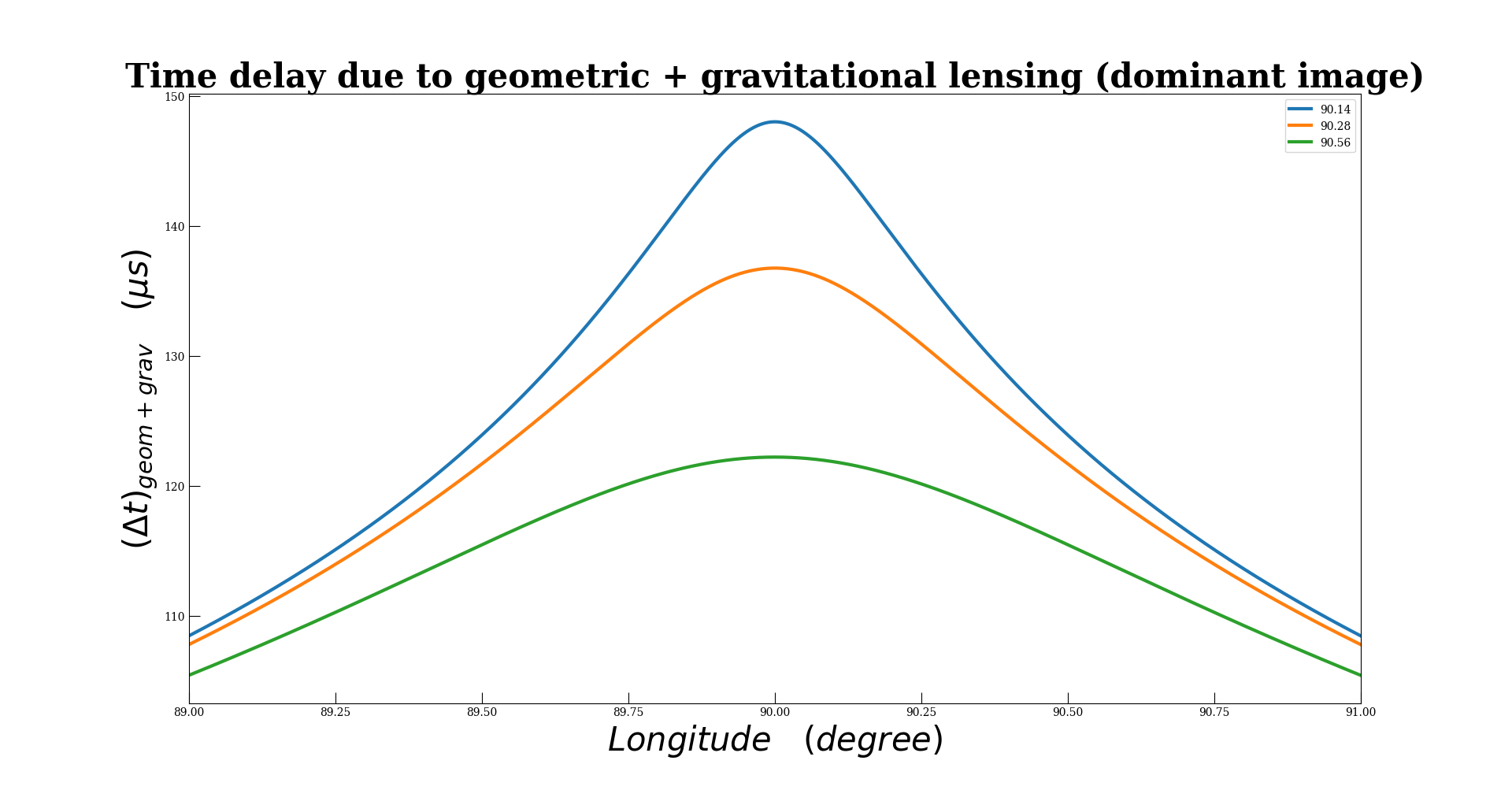}\hfill
    \includegraphics[width=.5\textwidth]{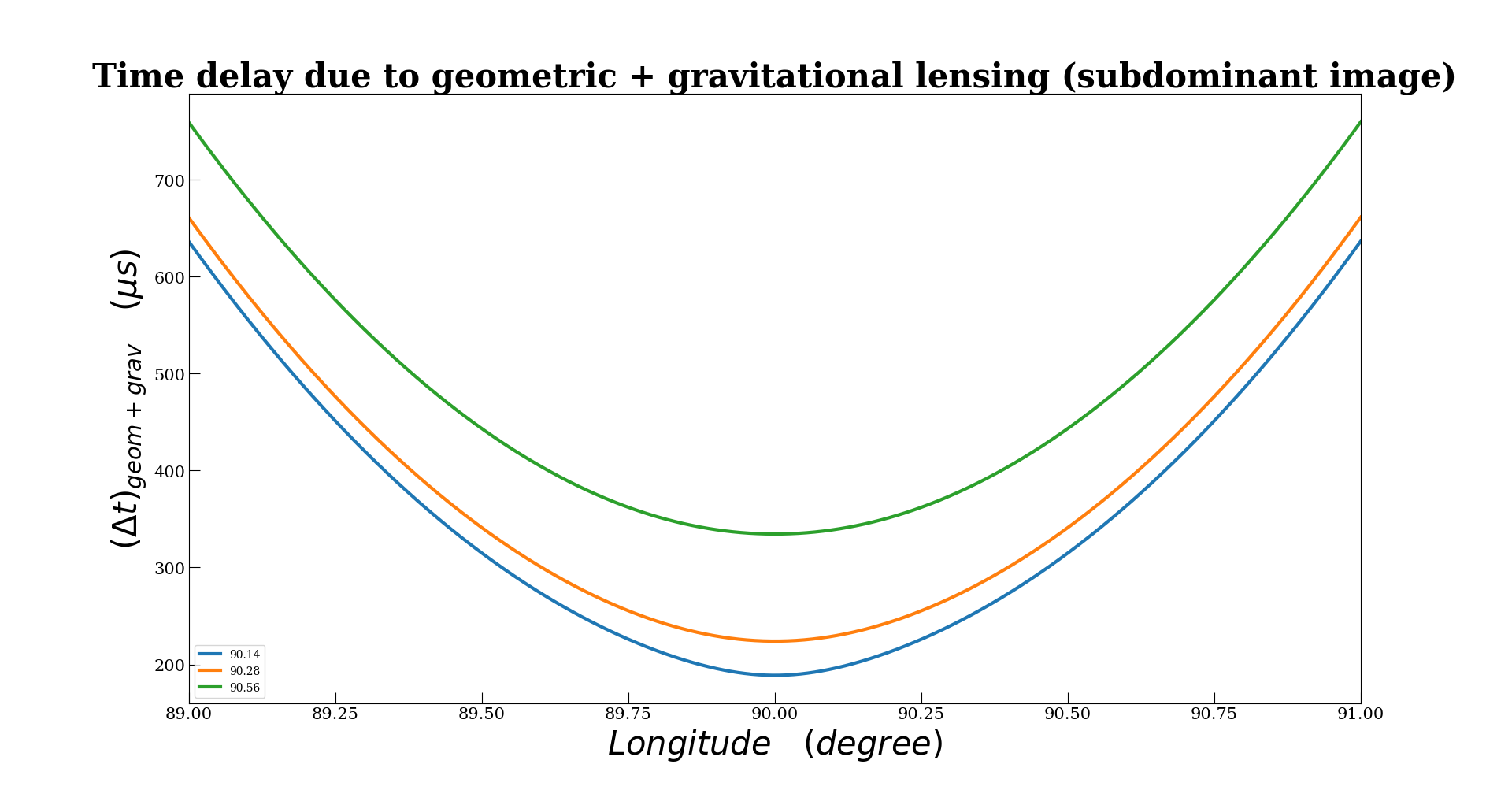}
    \newline    
    \includegraphics[width=.5\textwidth]{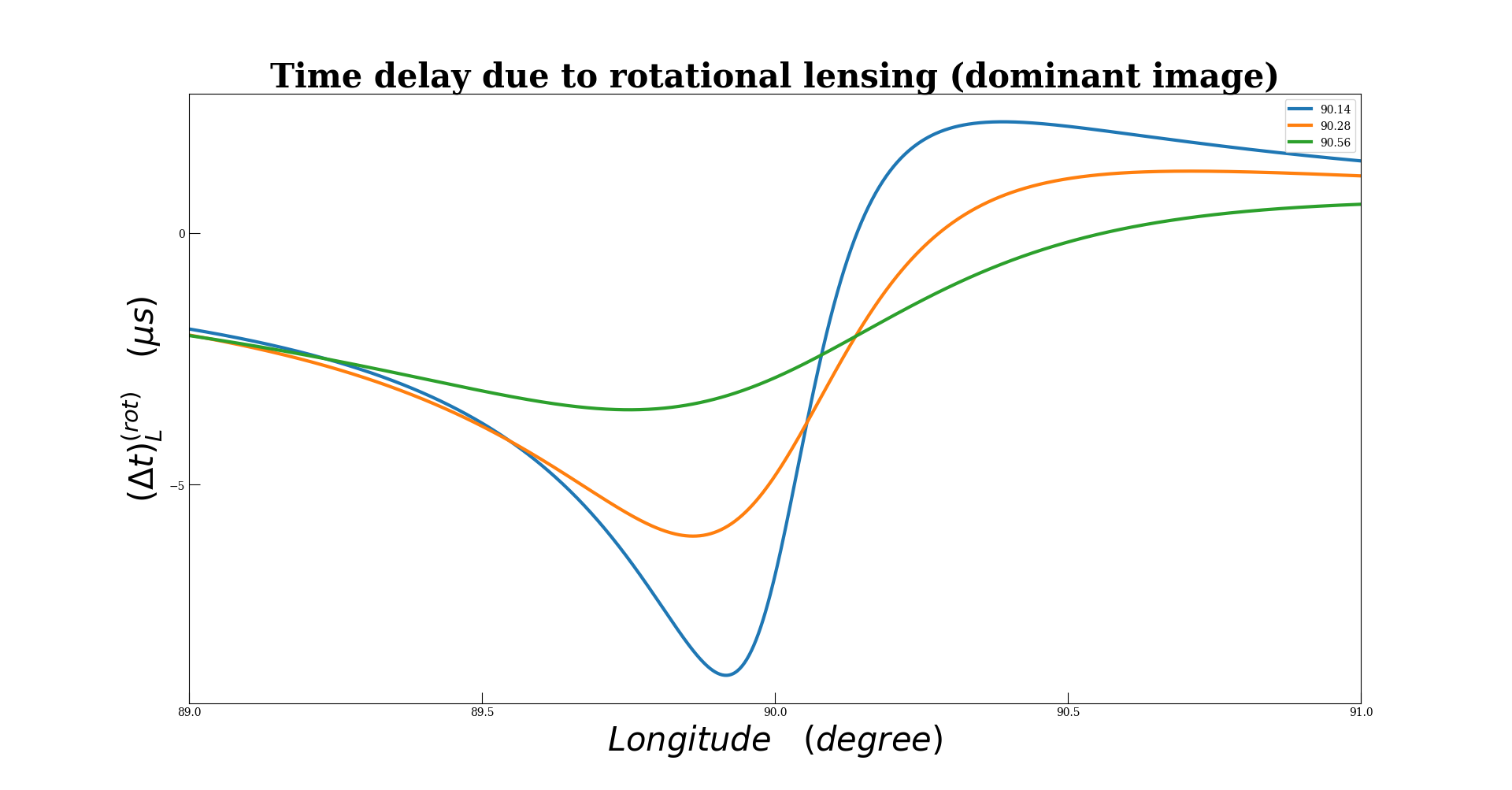}\hfill
    \includegraphics[width=.5\textwidth]{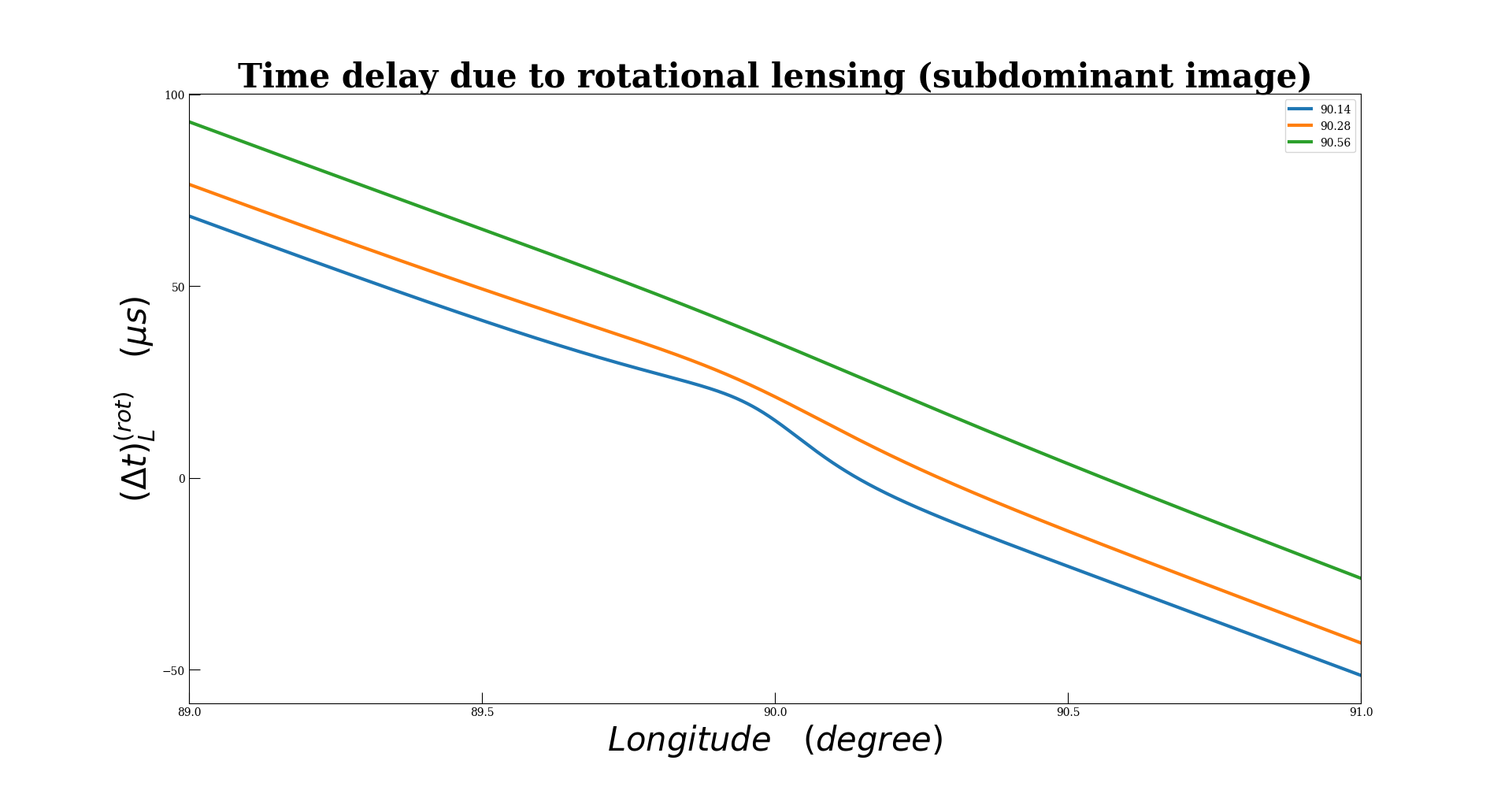}
    \newline
    \includegraphics[width=.5\textwidth]{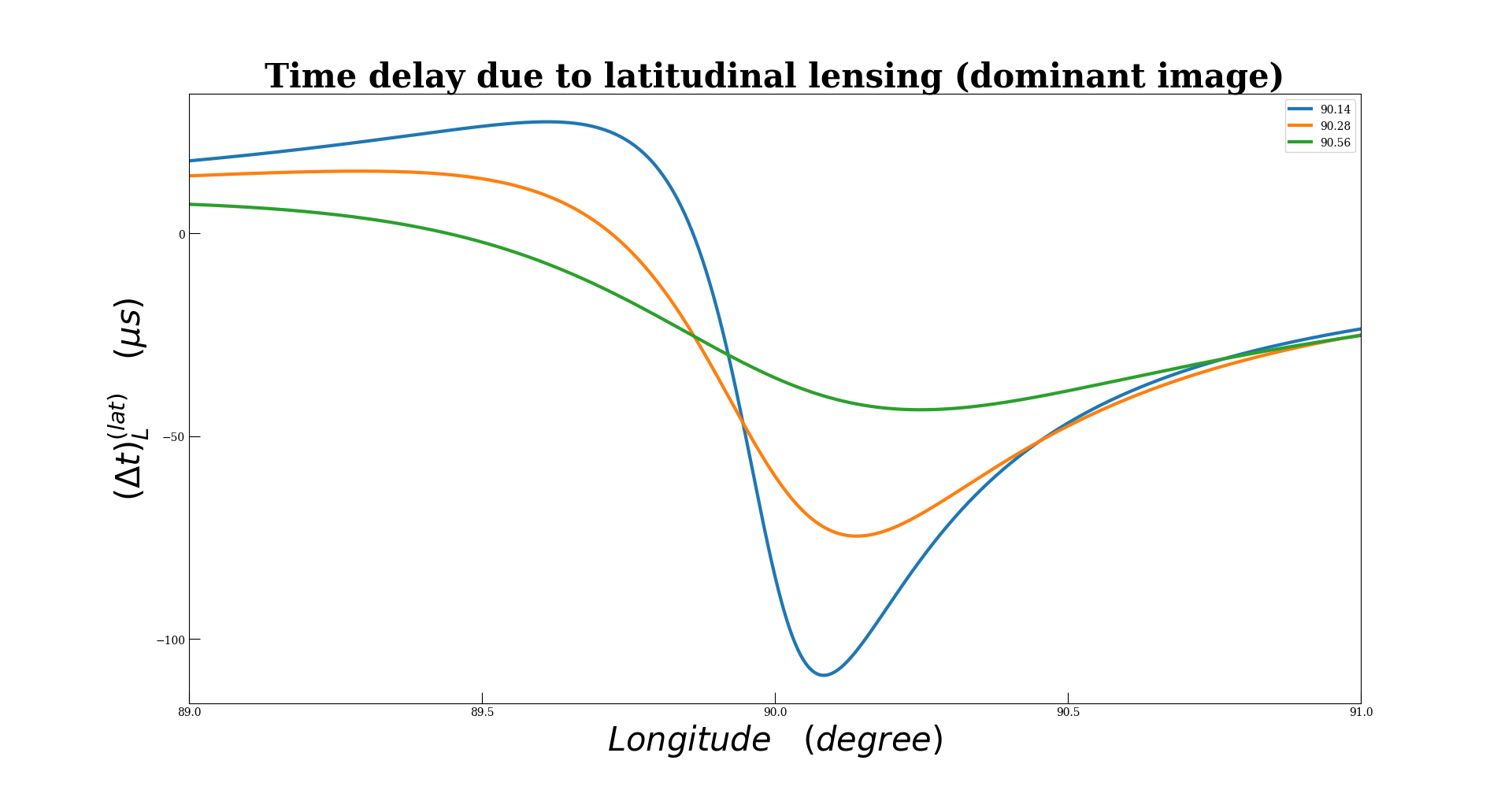}\hfill
    \includegraphics[width=.5\textwidth]{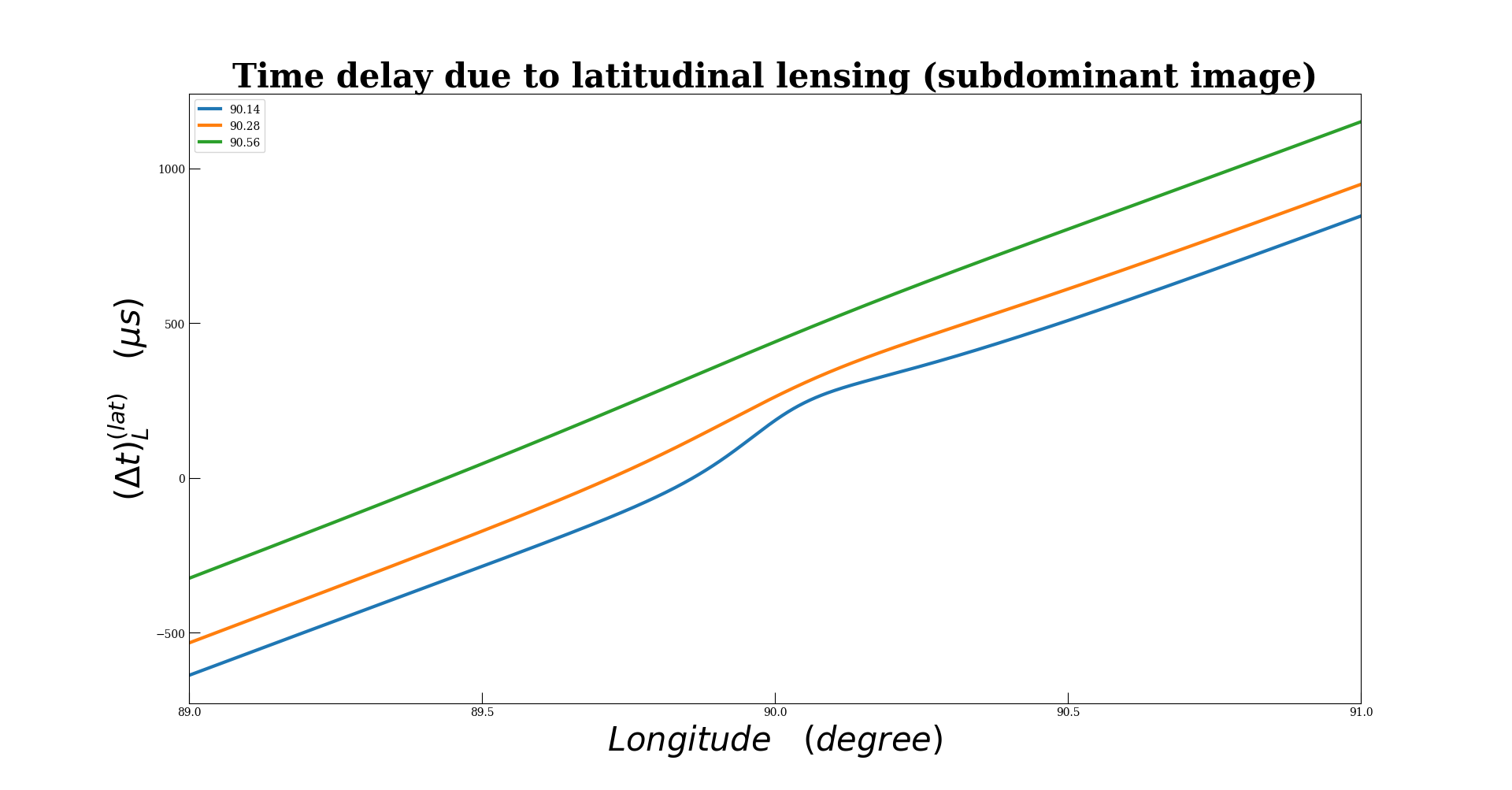}\hfill
    \caption{Amplification factor, combined geometric and gravitational delay, rotational lensing and latitudinal lensing delay plots for both dominant (left) and subdominant (right) images of `A' at superior conjunction \citep{McLaughlin2004}}
\end{figure}
PERISTOLE offers the user the ability to plot upto 6 different graphs related to a binary pulsar-body system such as the magnification factor of the pulse due to gravitational lensing, the geometric, Shapiro and the combine geometric and gravitational delay, rotational and latitudinal lensing delay.

\begin{itemize}
    \item Amplification/Magnification factor \begin{equation}
    A=\frac{\left(\frac{R}{R_E}\right)^2+2}{\left(\frac{R}{R_E}\right) \sqrt{\left(\frac{R}{R_E}\right)^2+4}}
    \end{equation}
    \item Geometric delay
    \begin{equation}
    (\Delta t)_{\mathrm{geometric}}=\frac{R_g}{c}\left(\frac{\Delta R_{\pm}}{R_E}\right)^2
    \end{equation}
    \item Shapiro delay
    \begin{equation}
    (\Delta t)_{\mathrm{gravitational}}=-\frac{R_g}{c} \ln \left[\frac{\sqrt{{r\sin i\sin\psi}^2+R_{\pm}^2}-r\sin i\sin\psi}{a\left(1-e^2\right)}\right]
    \end{equation}
    \item Rotational lensing delay
    \begin{equation}
    (\Delta t)_{rotational}=-\frac{\Delta R_{\pm}}{R} \frac{r}{a_{\|}} \frac{\sin \eta \cos \psi-\cos i \cos \eta \sin \psi}{\Omega_p \sin \zeta}
    \end{equation}
    \item Latitudinal lensing delay
    \begin{equation}
    (\Delta t)_{latitudinal}=\frac{\Delta R_{\pm}}{R} \frac{r}{a_{\|}} \frac{\cos \eta \cos \psi+\cos i \sin \eta \sin \psi}{\Omega_p \sin \zeta \tan \chi_0}
    \end{equation}
\end{itemize}
where $i$ is the orbital phase, $a$ is the orbital semimajor axis, $R_E$ is the Einstein Radius and $e$ is the eccentricity of the system, for an in-depth study of the formulae, please refer to \cite{Rafikov2005} and \cite{Rafikov2006} and for more information about the user defined parameters, please refer to the \href{http://peristole.readthedocs.io/}{documentation}. To demonstrate the capabilities of PERISTOLE, we recreate the amplification, combined geometric and Shapiro delays, rotational lensing delay and latitudinal lensing delay plots for both the dominant and sub-dominant images of PSR J0737-3039A as a function of the orbital phase from \cite{Rafikov2006}.

\section*{Installation, Documentation, and Future}

PERISTOLE is made available to the open source community to encourage functionalities, QoL improvements and better science. It can be found on Github\footnote{\url{https://github.com/centarsirius/peristole}} and is archived in Zenodo \cite{peridoi}. To aid seamless installation of the package, it is available on PyPi\footnote{\url{https://pypi.org/project/peristole/}} and can be installed via the pip command. The Github page also contains a jupyter notebook hosted on google colab to allow users to perform low volume analysis or quickly check out the package before installing. PERISTOLE can output publication-ready plots and further documentation is available at \url{http://peristole.readthedocs.io}. We hope to add more functionalities in the future such as analyzing the effects of the system on its immediate environment, reduce the number of parameters required to graph the time delays and automatically import system parameters from online archives like MAST and SIMBAD based on the system identifier.

The authors would like to thank the organizers of the Code/Astro 2022 workshop\footnote{\url{https://semaphorep.github.io/codeastro/}} for imparting the knowledge required to develop astronomy related software and the Code/Astro community for their constant support throughout the development of this package.

%

\vspace{5mm}

\software{numpy \citep{numpy}, matplotlib \citep{plt}}

\bibliography{sample631}{}

\begin{thebibliography}{}
\expandafter\ifx\csname natexlab\endcsname\relax\def\natexlab#1{#1}\fi
\providecommand{\url}[1]{\href{#1}{#1}}
\providecommand{\dodoi}[1]{doi:~\href{http://doi.org/#1}{\nolinkurl{#1}}}
\providecommand{\doeprint}[1]{\href{http://ascl.net/#1}{\nolinkurl{http://ascl.net/#1}}}
\providecommand{\doarXiv}[1]{\href{https://arxiv.org/abs/#1}{\nolinkurl{https://arxiv.org/abs/#1}}}

\bibitem[{{Damour} \& {Taylor}(1992)}]{Taylor1992}
{Damour}, T., \& {Taylor}, J.~H. 1992, \prd, 45, 1840,
  \dodoi{10.1103/PhysRevD.45.1840}

\bibitem[{Harris {et~al.}(2020)Harris, Millman, van~der Walt, Gommers,
  Virtanen, Cournapeau, Wieser, Taylor, Berg, Smith, Kern, Picus, Hoyer, van
  Kerkwijk, Brett, Haldane, del R{\'{i}}o, Wiebe, Peterson,
  G{\'{e}}rard-Marchant, Sheppard, Reddy, Weckesser, Abbasi, Gohlke, \&
  Oliphant}]{numpy}
Harris, C.~R., Millman, K.~J., van~der Walt, S.~J., {et~al.} 2020, Nature, 585,
  357, \dodoi{10.1038/s41586-020-2649-2}

\bibitem[{Hunter(2007)}]{plt}
Hunter, J.~D. 2007, Computing in Science \& Engineering, 9, 90,
  \dodoi{10.1109/MCSE.2007.55}

\bibitem[{{Komesaroff}(1970)}]{Komesaroff1970}
{Komesaroff}, M.~M. 1970, \nat, 225, 612, \dodoi{10.1038/225612a0}

\bibitem[{{Lai} \& {Rafikov}(2005)}]{Rafikov2005}
{Lai}, D., \& {Rafikov}, R.~R. 2005, \apjl, 621, L41, \dodoi{10.1086/429146}

\bibitem[{{Lorimer}(2005)}]{Lorimer2005}
{Lorimer}, D.~R. 2005, Living Reviews in Relativity, 8, 7,
  \dodoi{10.12942/lrr-2005-7}

\bibitem[{{McLaughlin} {et~al.}(2004){McLaughlin}, {Lyne}, {Lorimer},
  {Possenti}, {Manchester}, {Camilo}, {Stairs}, {Kramer}, {Burgay}, {D'Amico},
  {Freire}, {Joshi}, \& {Bhat}}]{McLaughlin2004}
{McLaughlin}, M.~A., {Lyne}, A.~G., {Lorimer}, D.~R., {et~al.} 2004, \apjl,
  616, L131, \dodoi{10.1086/426813}

\bibitem[{{Rafikov} \& {Lai}(2006)}]{Rafikov2006}
{Rafikov}, R.~R., \& {Lai}, D. 2006, \apj, 641, 438, \dodoi{10.1086/500346}

\bibitem[{Venkatesh \& Pundir(2022)}]{peridoi}
Venkatesh, S., \& Pundir, G. 2022, centarsirius/peristole: v1.1.0,  Zenodo,
  \dodoi{10.5281/zenodo.6744000}

\end{thebibliography}
\bibliographystyle{aasjournal}



\end{document}